\documentstyle[multicol,aps,prb,epsf]{revtex}
\begin{document}
\draft

\title{Electronic Structure of Sodium Cobalt Oxide:
Comparing Mono- and Bilayer-hydrate}
\author{Ryotaro Arita$^{1,2}$}
\address{$^1$Max Planck Institute for Solid State Research,
Heisenbergstr. 1, Stuttgart 70569, Germany }
\address{$^2$Department of Physics, University of Tokyo, Hongo,
Tokyo 113-0033, Japan}
\date{\today}
\maketitle

\begin{abstract}
To shed new light on the mechanism of superconductivity
in sodium cobalt oxide bilayer-hydrate (BLH), we perform
a density functional calculation with full structure 
optimization for BLH and its related nonsuperconducting phase, 
monolayer hydrate (MLH). We find that these hydrates have
similar band structures, but a notable difference can
be seen in the $a_{1g}$ band around the Fermi level.
While its dispersion in the $z$ direction is 
negligibly small for BLH, it is of the order of 0.1 eV for MLH. 
This result implies that the three dimensional feature of
the $a_{1g}$ band may be the origin for the absence 
of superconductivity in MLH. 
\end{abstract}
\medskip

\begin{multicols}{2}
\narrowtext

Layered sodium cobalt oxide bilayer-hydrate (BLH)
is now attracting mounting attention due
to the recent discovery of its superconductivity\cite{Takada03}.
One of the interesting characteristic feature of 
this superconducting phase is that it is derived through 
a soft-chemical process for the parent 
material (Na$_{0.7}$CoO$_2$). Namely, a part of the Na ions 
located between the CoO$_2$ layers are extracted and 
H$_2$O molecules are inserted for this region.
Here, the role of H$_2$O 
is of great interest from both a chemical and a physical 
point of view,
since it can be a key to understand the pairing mechanism, 
and furthermore, it may provide some 
guiding principles to synthesize other novel superconductors 
in layered transition metal oxides. 

It has been proposed from the beginning\cite{Takada03} 
that the main effect of 
inserting H$_2$O molecules is expanding the interlayer distance ($d$)
of the CoO$_2$ layers. Namely, while 
$d=$5.4 \AA\ for the parent material, that of 
superconducting BLH is expanded to
$d=$9.8 \AA. Electronic structure calculations within the density 
functional theory\cite{Johannes04,Marianetti04} have indeed shown
that the electronic structure of BLH is highly two-dimensional,
whereas the $c$-axis dispersion of the band structure
as well as the bonding-antibonding band splittings due to
the interlayer couplings
are not negligible for unhydrated Na$_x$CoO$_2$. 
In fact, the two-dimensional electronic structure is favorable 
for unconventional superconductivity where 
the pairing interaction has a characteristic structure in the 
momentum space\cite{Arita99,Monthoux99}.

On the other hand, to provide a deep insight into the role of the
H$_2$O molecules, Takada {\it et al.} synthesized monolayer-hydrate 
(MLH) by partial extraction of H$_2$O from BLH, and studied
the magnetic properties\cite{Takada03b,Sakurai03}. 
One may expect that MLH is also superconducting
because $d=$6.9 \AA, which
is larger than that of La$_2$CuO$_4$ (=6.6 \AA). However, 
they found that the superconductivity is completely suppressed, 
while the carrier density in the CoO$_2$ layer is expected to 
be unchanged because the magnetic susceptibilities of MLH 
and BLH have similar values above 20K. Thus Takada {\it et al.}
concluded that the H$_2$O molecules in BLH may play an 
important role for the superconductivity besides the separation 
of the CO$_2$ layers. In fact, in contrast to MLH where 
the H$_2$O molecules 
and the Na ions are placed on the same plane, in BLH,
the H$_2$O molecules are 
placed between the CoO$_2$ layers and the Na ions. Therefore,
they proposed that the H$_2$O molecules in BLH may shield the 
(random) Coulomb potential of the Na ions.

While intensive first-principles calculations for BLH have been done 
up to present\cite{Lee04,Zhang04,Zou04,Johannes04b,Zhang04b}, 
those for MLH have yet to be carried out. 
The purpose of the present study is 
to investigate why BLH is superconducting but MLH is not
by comparing the electronic structure of BLH and MLH,
and to obtain some hints to understand 
the mechanism of superconductivity in BLH. 
We found that BLH and MLH have
similar band structures, but a notable difference can
be seen in the $a_{1g}$ band around the Fermi level
in that while its dispersion in the $z$ direction is 
negligibly small for BLH, it is of the order of 0.1 eV for MLH.

Although various theoretical studies have been performed 
for the superconductivity in BLH, 
the mechanism is still controversial.
Especially, clarifying which bands among the $t_{2g}$ bands
around the Fermi level
dominate the superconductivity is of great interest. 
A variety of scenarios have been proposed. In fact, while there 
have been several works which claim that the $a_{1g}$ bands 
are important
\cite{Baskaran03,Honerkamp03,Tanaka04,Shastry03,Singh03,Ogata03,PALee03} or both $a_{1g}$ and $e'_g$ should be 
considered\cite{Koshibae03,Mochizuki04},
the present author and his collaborators have
proposed that the pocket-like Fermi surfaces of the $e'_g$
bands are essential\cite{Kuroki04}. 
Here, the advantage of first principles calculations
is that it can give some hints on which band is important 
and should be taken into account for the model calculation.
The present result that the main difference between 
BLH and MLH is the dispersion of the $a_{1g}$ bands 
suggests that the $a_{1g}$ band should not be ignored
when we study superconductivity in BLH,
and the three dimensional feature of
the $a_{1g}$ band may be the origin for the absence 
of superconductivity in MLH. 

In the present study, 
we perform a first-principles band structure calculation within the
framework of the generalized gradient approximation (GGA) based 
on the density functional theory. In the calculations, the 
exchange-correlation functional introduced by
Perdew, Burke, and Wang\cite{Perdew96} is adopted with the 
ultra-soft pseudopotential\cite{Vanderbilt90,Laasonen93} 
in a separable form. The wave functions are expanded by 
plane waves up to a cut-off energy of 30.25 Ry, and 
$k$ point meshes of $8\times 8\times 2$ are used.
The atomic configurations are optimized to minimize the total 
energy with the conjugate gradient scheme\cite{Yamauchi96}.
We take a $\sqrt{3}a\times\sqrt{3}a\times c$ supercell,
in which six cobalt atoms are included.
During the optimization, the lattice constant $a$ and $c(=2d)$ 
are fixed to the experimental values, i.e., $a$=2.8169 (2.8345) \AA\
and $c$=19.645 (13.842) \AA\ 
for BLH (MLH)\cite{Takada03b,Jorgensen03}.
While a variety of charge and spin orderings are observed 
in Na$_x$CoO$_2$ for $x\geq$0.5, such orders have not been observed 
in BLH and MLH ($x \sim$ 0.3). Therefore, we performed a density
functional calculation for paramagnetic states without assuming 
any charge orderings for the initial state of the self-consistent
calculation.

In this paper, we focus on the electronic structure
for Na$_{1/3}$CoO$_2$(H$_2$O)$_{2/3}$  (BLH) and 
Na$_{1/3}$CoO$_2$(H$_2$O)$_{1/3}$ (MLH). On the other hand,
recently, it was reported that there are not only Na 
ions but also oxonium ions ((H$_3$O)$^+$) between the 
CoO$_2$ layers, 
so that the formal valence Co should be +3.4, which is 
lower than +3.65 as estimated from the 
Na content\cite{Takada04}.
While this means that the Fermi level locates slightly higher 
than that of the present calculation, for simplicity, we neglect 
the oxonium ions. However, as is discussed below,
we expect that the shift of the Fermi level does not
change the qualitative tendency of the present result.

\begin{figure}
\begin{center}
\leavevmode\epsfysize=70mm \epsfbox{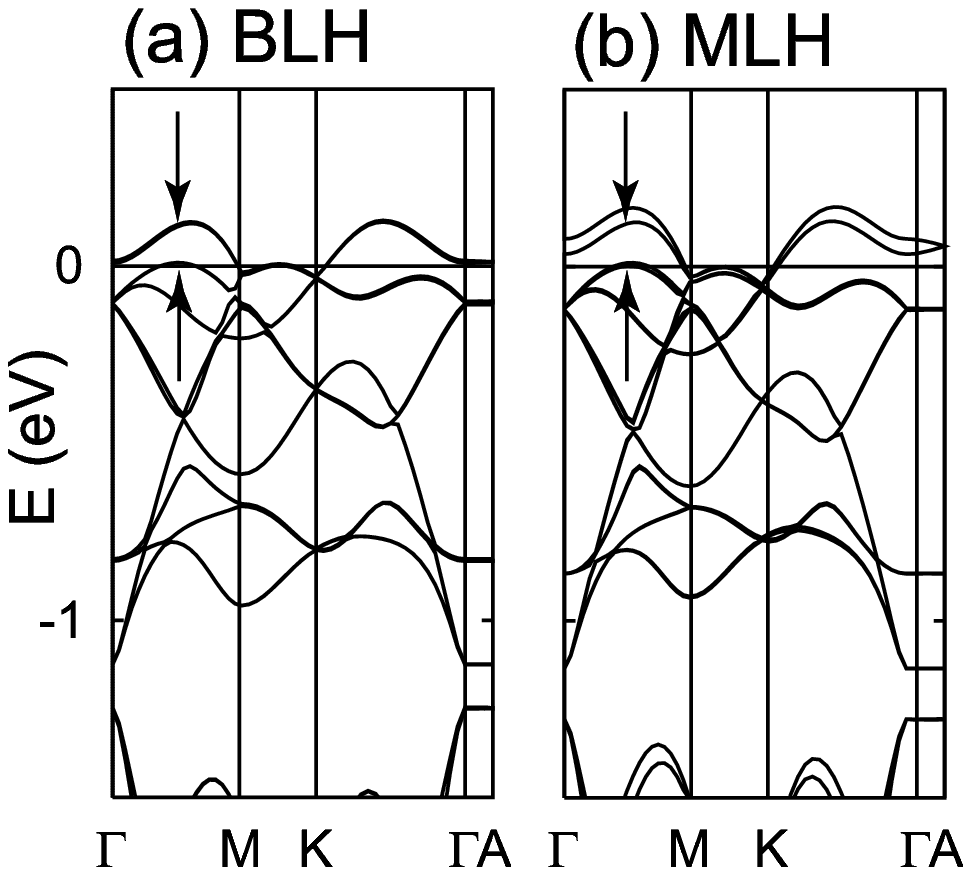}
\caption{The band structure of CoO$_2$, for which the effect 
of the Na and the H$_2$O molecules is
represented by increasing the number of electrons in the system. 
The energies are measured from the Fermi level.
The lattice constants are set 
to be those of (a) BLH and (b) MLH. A small band splitting can 
be seen for the $a_{1g}$ bands just above the Fermi level.
Solid arrows point onto the $a_{1g}$ and $e'_{g}$ band
at the $k$-points for which the Bloch wave 
functions are shown in Fig.\ref{fig2}.
}
\label{fig1}
\end{center}
\end{figure}

Before the actual calculation for
Na$_{1/3}$CoO$_2$(H$_2$O)$_{2/3}$ and
Na$_{1/3}$CoO$_2$(H$_2$O)$_{1/3}$, 
we first introduced a reference system where 
the Na ions and the H$_2$O molecules is represented
by a corresponding electron doping into the
CoO$_2$ layers. Namely, we calculated the electronic 
structure of the CoO$_2$ layers with an increased number of 
electrons and a uniform positive background charge 
to make the system charge-neutral.
This reference system represents a situation where
the H$_2$O molecules completely shield the potentials of 
the Na ions.

In Fig. \ref{fig1}(a) and (b), we show the band structure
of the reference system
where the lattice constants are set to those 
of BLH\cite{Jorgensen03} and MLH\cite{Takada03b}, respectively. 
Here, the eighteen bands comprising the $t_{2g}$ manifold 
are split into six $a_{1g}$ and twelve $e'_g$ bands.
Note that only nine bands would be distinguishable if 
the two layers in the unit cell did not interact. 
We can see in Fig. \ref{fig1}
that while the band dispersion in (a) and (b) are 
almost identical to each other, for (b) a small band splitting
can be seen in the $a_{1g}$ bands just above the Fermi level.

\begin{figure}
\begin{center}
\leavevmode\epsfysize=50mm \epsfbox{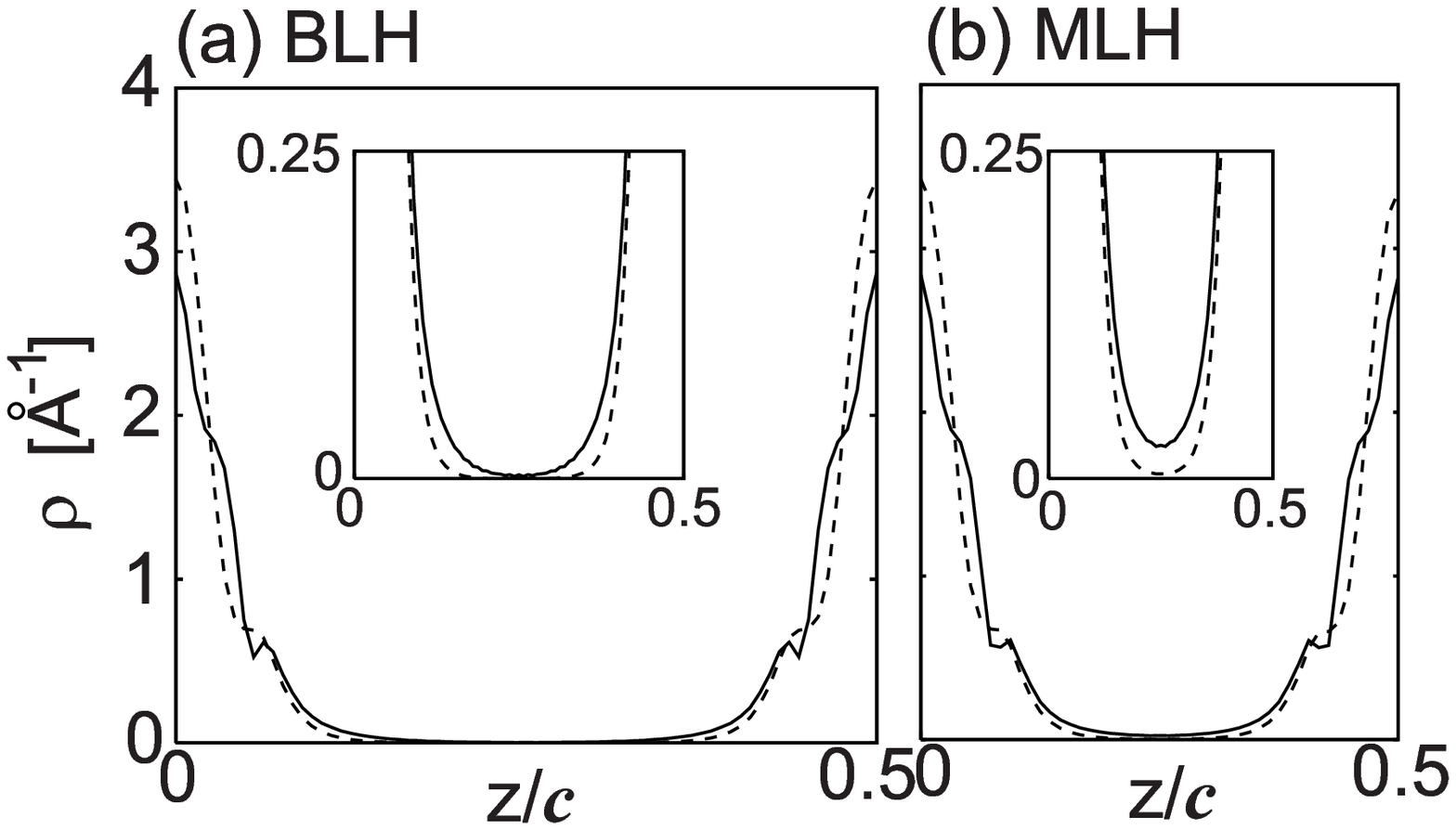}
\caption{
Squared absolute value of the Bloch wave function
at the $k$-points indicated in Fig. \ref{fig1} 
for the bonding $a_{1g}$ (solid line)
and $e'_{g}$ band (dotted line).
Only lower half of the unit cell is shown and 
the Co planes reside around $z=0$ and $0.5c$. 
The inset is an enlargement of $\rho \leq 0.25$.
}
\label{fig2}
\end{center}
\end{figure}

In Fig. \ref{fig2}, we plot 
$\rho(z)=\int |\phi_{\bf k}({\bf r})|^2 dxdy$, where
$\phi_{\bf k}$ is the Bloch wave function
of the $k$-point indicated with arrows in Fig. \ref{fig1}
on the $a_{1g}$ and $e'_{g}$. 
Here, only the lower half of the unit cell is shown,
and the Co planes reside around $z=0$ and $0.5c$.
We can see that the main amplitude of $\rho(z)$ is
localized within the Co layers. For BLH, the peaks
on the Co planes are well separated for both 
$a_{1g}$ and $e'_{g}$. For MLH,
on the other hand, the $a_{1g}$ state has finite amplitude 
there whereas the $e'_{g}$ state vanishes $z\sim 0.25c$.

These results suggest that the assumption that the effect 
of the Na ions and the H$_2$O molecules is irrelevant can
be invalid for the $a_{1g}$ states in MLH. Especially,
the effect of the random potential of Na can be serious
for these states. In other words, it is suggested that
(i) for unhydrated Na$_x$CoO$_2$\cite{Johannes04,Marianetti04},
the $c$-axis dispersion in the band structure or
the interlayer couplings are not negligible for both 
$e'_{g}$ and $a_{1g}$ orbitals because 
$d$ is not sufficiently large; (ii) for MLH, $d$ is large 
enough to suppress the interlayer transfers or 
couplings for $e'_{g}$ but not for $a_{1g}$ orbitals,
because the wave functions of the latter 
extends to the $c$-axis;
(iii) for BLH, $d$ is sufficiently large so that the whole
$t_{2g}$ manifold is highly two-dimensional.
It is expected that this qualitative tendency would not 
change even if we increased the amount of electron doping\cite{note1}.

\begin{figure}
\begin{center}
\leavevmode\epsfysize=50mm \epsfbox{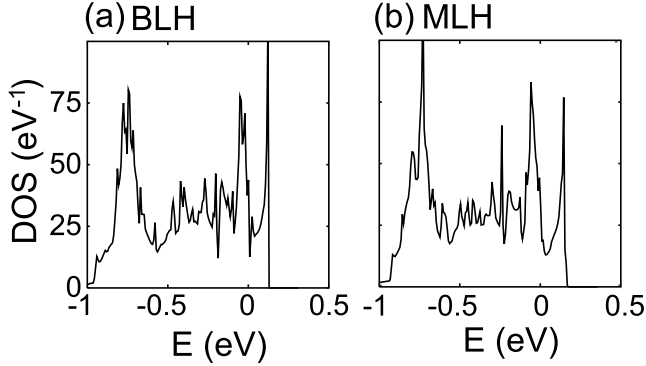}
\caption{
Density of states for the band structures
of Fig.\ref{fig1}.
}
\label{fig3}
\end{center}
\end{figure}

In Fig. \ref{fig3}, we plot the density of states for  
Fig. \ref{fig1}. We can see that (a) and (b) are 
similar to each other. Especially, both have peaks 
just below the Fermi level, which are due to 
the van Hove singularity in the $e'_g$ bands\cite{Kuroki04}. 
On the other hand, the divergence at the upper band edge is weak 
for (b) due to the band splitting of the $a_{1g}$ bands.

\begin{figure}
\begin{center}
\leavevmode\epsfysize=50mm \epsfbox{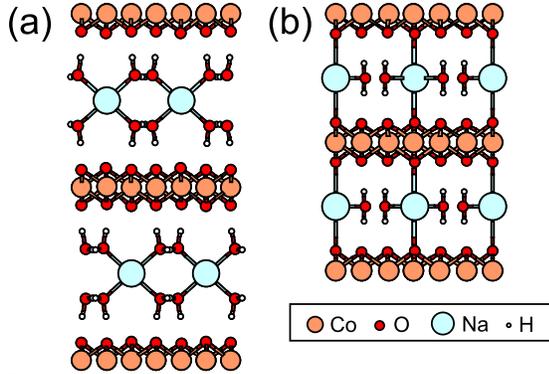}
\caption{The optimized atomic configuration of 
(a) Na$_{1/3}$CoO$_2$(H$_2$O)$_{2/3}$ and 
(b) Na$_{1/3}$CoO$_2$(H$_2$O)$_{1/3}$.
}
\label{fig4}
\end{center}
\end{figure}

Next, let us discuss the actual calculation for
BLH and MLH, i.e., Na$_{1/3}$CoO$_2$(H$_2$O)$_{2/3}$ and 
Na$_{1/3}$CoO$_2$(H$_2$O)$_{1/3}$.
For the former, starting with the atomic configuration 
by Jorgensen {\it et al.} (Fig. 8a of Ref.\onlinecite{Jorgensen03}), 
we performed a structure optimization. In Fig. \ref{fig4}(a), 
we show the side view of the optimized structure.
As for the initial atomic configuration of the latter, 
we followed Takada {\it et al}.\cite{Takada03b} 
As is stressed in Ref. \cite{Takada03b},
it should be noted that the positions of the Na ions in MLH
are different from those in BLH. Namely, while for BLH the Na 
atoms locate near the center of the trigonal prisms formed by 
the facing oxygen atoms of adjacent CoO$_2$ layers,
they locate the midpoints 
of the prism edge for MLH (Fig. 4 in Ref\cite{Takada03b}).
Since the atomic position of the H$_2$O 
molecules are not given in Ref\cite{Takada03b}, 
we started with several patterns of initial configurations,
placing the oxygen atoms of the H$_2$O molecules
on the same plane as the Na atoms. While we show the 
result for the most stable structure (Fig. \ref{fig4}(b)),
the following argument does not depend on the
detail of the atomic configuration of H$_2$O atoms\cite{note3}. 

\begin{figure}
\begin{center}
\leavevmode\epsfysize=70mm \epsfbox{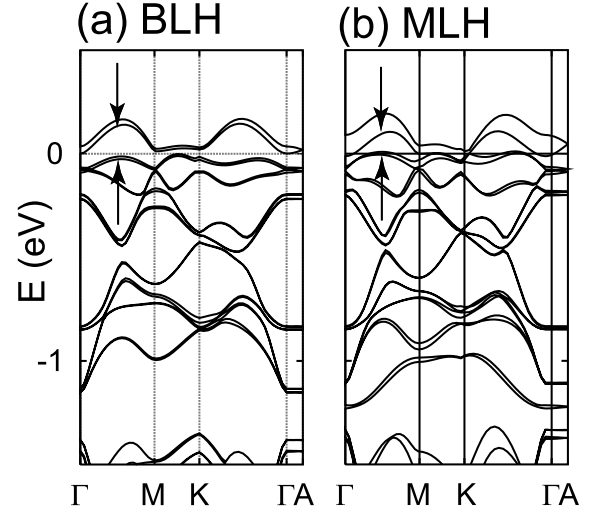}
\caption{
A plot similar to Fig. \ref{fig1} of
the atomic configuration of Fig. \ref{fig4}.
}
\label{fig5}
\end{center}
\end{figure}

In Fig. \ref{fig5}, we show the resulting band structure.
The difference between Fig. \ref{fig1} and Fig. \ref{fig5}
is whether the intercalated hydrate group is actually 
considered or not.  By comparing Fig. \ref{fig1} 
and Fig. \ref{fig5}, we can evaluate the effect of the 
potential of Na ions, the screening by the H$_2$O 
molecules, etc. We can see that the effect for the $e'_g$ 
bands in BLH and MLH is similar to each other. 
Especially, for MLH, the band splitting in the $e'_g$ bands 
just below the Fermi level is negligibly small, which
contrasts with the case of unhydrated Na$_x$CoO$_2$, for 
which the band splitting is
as large as 0.1 eV\cite{Johannes04,Marianetti04}.
This result implies that inserting single layer of H$_2$O 
molecules is sufficient to suppress the interlayer couplings 
for the $e'_g$ bands.

On the other hand, the $c$-axis dispersion and the 
band splitting of the $a_{1g}$ bands
is of the order of 0.1 eV for MLH, while those for BLH 
are negligibly small. 
This suggests that inserting a single layer of H$_2$O molecules 
is not sufficient to separate the $a_{1g}$ states in each 
CoO$_2$ layer since they extend along the $c$-axis.
Therefore, we expect for MLH that while the $e'_g$ bands are
rather insensitive to the random potential of the Na ions, the
$a_{1g}$ bands feel them seriously.

\begin{figure}
\begin{center}
\leavevmode\epsfysize=50mm \epsfbox{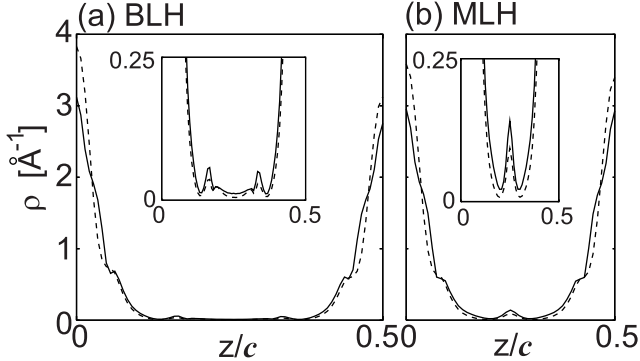}
\caption{
Same as Fig. \ref{fig2}, but now for the atomic 
configuration of Fig. \ref{fig4}.
}
\label{fig6}
\end{center}
\end{figure}

In Fig. \ref{fig6}, we plot $\rho$ for the $k$-points 
on the bonding $a_{1g}$ and $e'_g$ band
which are indicated in Fig. \ref{fig5}. 
The $c$-axis dispersion is predominantly determined by 
the direct overlap between the tails of the adjacent
main peaks rather than the amplitude on the H2O molecules 
(minor middle peak). We can see that the overlap is finite
for the $a_{1g}$ band of MLH but negligibly small for the
other cases. (Note that for the $e'_g$ band of MLH,
$\rho$ vanishes between the main peak and the small peak
around the oxygen site in the H$_2$O molecule.) 

\begin{figure}
\begin{center}
\leavevmode\epsfysize=50mm \epsfbox{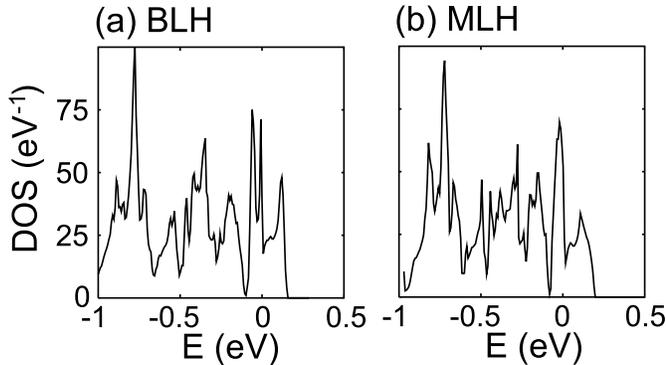}
\caption{
A plot similar to Fig. \ref{fig3}, for the atomic configuration
of Fig. \ref{fig4}.
}
\label{fig7}
\end{center}
\end{figure}

In Fig. \ref{fig7}, we show the density of states
for the band structure shown in Fig. \ref{fig5}. 
While the overall qualitative feature of Fig. \ref{fig7} (a)
and (b) is similar to each other, the divergence
at the upper band-edge for BLH is completely smeared out
for MLH.

In conclusion, we found that BLH and MLH have quite
similar band structures, but there is a notable difference
in the dispersion of the $a_{1g}$ band around the Fermi level. 
Namely, it is negligibly small in the $z$ direction for the former, 
it is an order of 0.1 eV for the latter. This suggests that 
although a single layer of H$_2$O molecules is sufficient
to suppress the interlayer coupling for the $e'_g$ bands,
it is not sufficient for the $a_{1g}$ bands. 

If we assume that the three dimensional feature of 
the $a_{1g}$ band can destroy superconductivity, 
we can understand 
why BLH is superconducting but MLH is not. In this sense, 
the present result suggests that the $a_{1g}$ bands should 
not be ignored when we study superconductivity
in Na$_x$CoO$_2 \cdot y$H$_2$O.

Recently, a systematic angle-resolved photoemission study has 
been performed for wide range of Na concentrations\cite{Yang}.
They concluded that the Fermi surface consists of the $a_{1g}$ band
for $x=$0.3, which suggests that the $a_{1g}$ band is important
for the superconductivity. This is consistent with the present
conclusion.

The GGA calculation was performed with TAPP (Tokyo ab-initio 
program package), for which we received technical advices from 
Y. Suwa.  We benefitted from fruitful discussions with
K. Kuroki. We would like to thank M.A. Korotin for providing us a 
program of the tetrahedron method and 
K. Held for his critical reading of the manuscript.
This research was supported by the Alexander von Humboldt Foundation.

\end{multicols}

\begin{references}
\bibitem{note1}This qualitative tendency should not depend on the choice of
the exchange-correlation potential. Even if we stepped forward 
to more elaborated calculations such as LDA+U or LDA+DMFT, it 
would not be changed.
\bibitem{note3}Here, for MLH (BLH), the distance between H in H$_2$O and 
O in CoO$_2$ plane is 1.72 (1.58-1.59) \AA, that between O 
in H$_2$O and Na is 2.28 (2.24) \AA, the bond length of Co-O is 
1.91-1.93 (1.90-1.93)\AA.
\end{references}
\end{document}